\documentclass[conference]{IEEEtran}
\IEEEoverridecommandlockouts
\usepackage{cite}
\usepackage{amsmath,amssymb,amsfonts}
\usepackage{algorithmic}
\usepackage{graphicx}
\usepackage{textcomp}
\usepackage{xcolor}
\usepackage{tabularx}
\usepackage{booktabs}
\def\BibTeX{{\rm B\kern-.05em{\sc i\kern-.025em b}\kern-.08em
    T\kern-.1667em\lower.7ex\hbox{E}\kern-.125emX}}
\begin{document}

\title{Strategic Cyber Defense via Reinforcement Learning-Guided Combinatorial Auctions}

\author{\IEEEauthorblockN{Mai Pham}
\IEEEauthorblockA{\textit{Thayer School of Engineering} \\
\textit{Dartmouth College}\\
Hanover, NH \\
mai.p.pham.th@dartmouth.edu}
\and
\IEEEauthorblockN{Vikrant Vaze}
\IEEEauthorblockA{\textit{Thayer School of Engineering} \\
\textit{Dartmouth College}\\
Hanover, NH\\
vikrant.s.vaze@dartmouth.edu}
\and
\IEEEauthorblockN{Peter Chin}
\IEEEauthorblockA{\textit{Thayer School of Engineering} \\
\textit{Dartmouth College}\\
Hanover, NH \\
peter.chin@dartmouth.edu}
}

\maketitle

\begin{abstract}
Cyber defense operations increasingly require long-term strategic planning under uncertainty and resource constraints. We propose a new use of combinatorial auctions for allocating defensive action bundles in a realistic cyber environment, using host-specific valuations derived from reinforcement learning (RL) Q-values. These Q-values encode long-term expected utility, allowing upstream planning. We train CAFormer, a differentiable Transformer-based auction mechanism, to produce allocations that are approximately incentive-compatible under misreporting. Rather than benchmarking against existing agents, we explore the qualitative and strategic properties of the learned mechanisms. Compared to oracle and heuristic allocations, our method achieves competitive revenue while offering robustness to misreporting. In addition, we find that allocation patterns correlate with adversarial and defensive activity, suggesting implicit alignment with operational priorities. Our results demonstrate the viability of auction-based planning in cyber defense and highlight the interpretability benefits of RL-derived value structures.
\end{abstract}

\begin{IEEEkeywords}
Cyber defense, strategic planning, mechanism design, differentiable optimization.
\end{IEEEkeywords}

\section{Introduction}
Modern cybersecurity operations can no longer rely solely on reactive measures, such as intrusion detection systems or manual incident response. As advanced persistent threats (APTs) and rapidly evolving malware campaigns continue to increase in sophistication, defenders must engage in strategic planning, taking the initiative in advance, to allocate limited defensive resources across an enterprise network \cite{zhu2015game, schlenker2017don}. These decisions must be made under adversarial pressure, where attackers exploit system complexity, resource misallocation, and latency in defense coordination.

Enterprise systems encompass a diverse set of assets: user endpoints, data servers, authentication services, IoT/OT devices (e.g., industrial controllers), and mission-critical applications such as SCADA or ERP platforms. These networks are modeled in high-fidelity simulations such as DARPA’s CAGE Challenge 2 (CC2) \cite{kiely2023cage} or observed in real-world datasets such as LANL’s enterprise event logs \cite{kent2015cybersecurity}. Protecting these heterogeneous assets requires selecting defensive actions—such as Analyze, Remove, or Restore—for a subset of hosts, constrained by human, operational, or computational limitations.

This notion of an “operational budget” is reflected in many abstract defense models. Stackelberg security games, for example, constrain defender strategies via probabilistic action budgets \cite{zhu2015game}, while robust Markov Decision Processes (MDPs) limit feasible action subsets to reflect bounded resources \cite{schlenker2017don}. Classical heuristics often address this by selecting the top k nodes based on static metrics such as risk scores or attack graph centrality \cite{lippmann2005attackgraph, yun2011cvss}. However, to our knowledge, no prior work explicitly formulates an upstream resource-constrained planning layer for the DARPA CC2 environment.

This paper introduces such a mechanism: a constrained auction-based planner that allocates defensive actions across hosts prior to episode execution. Guided by valuation signals derived from reinforcement learning agents, our approach reframes cyber planning as an economic coordination problem. Auctions are particularly well-suited to this task because they provide a principled, interpretable framework for allocating limited resources among self-interested, heterogeneous agents. Although underexplored in traditional cybersecurity settings, auctions have been applied effectively to related domains such as resource allocation and pricing in cloud environments \cite{zhang2018auction} and mitigation of denial-of-service (DoS) attacks \cite{zhang2018auction}, suggesting a broader applicability. Our abstraction enables modular, scalable, and strategic cyber planning, especially important when resource decisions must be made in advance, before threats fully materialize.

This upstream perspective differs from most existing work, which emphasizes reactive or tactical decision making. Game-theoretic approaches \cite{zhu2015game, alpcan2010network} model attacker-defender interactions with strong assumptions about observability and centralized control. Attack graph-based techniques \cite{sheyner2002automated, lippmann2005attackgraph} use structural analysis to identify vulnerable nodes but struggle to scale or adapt dynamically. MDPs and their extensions, including robust MDPs \cite{schlenker2017don}, Partially Observable MDPs (POMDPs) \cite{durkota2015optimal}, and reinforcement learning \cite{nguyen2023survey, li2021survey}, support sequential planning, but generally operate at the execution level and assume centralized planning.

\begin{figure}[htbp]
\centerline{\includegraphics[width=0.9\linewidth]{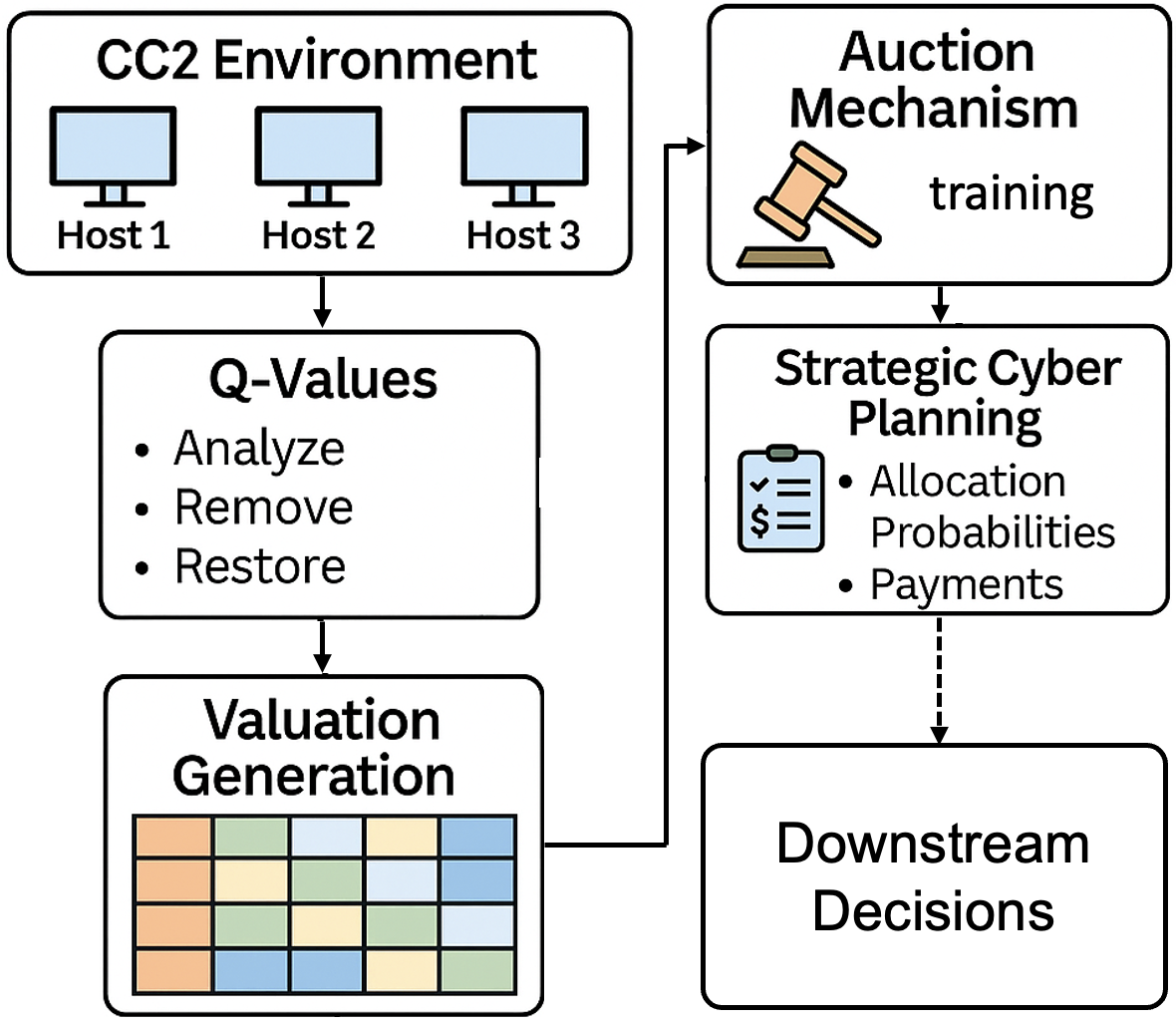}}
\caption{Overview of our cyber planning framework. Hosts in the DARPA CAGE Challenge 2 environment (e.g., user machines, servers) are modeled individually. Reinforcement learning agents generate Q-values for each possible defensive action—such as Analyze, Remove, or Restore—based on host behavior and environment context. These values are transformed into private valuations over action bundles. A trained auction mechanism then allocates limited defensive resources by computing probabilistic assignments and payments. The resulting strategic plan informs upstream resource allocation before any episode execution.}
\label{fig:pipeline}
\end{figure}

We propose a fundamentally different approach: modeling strategic cyber planning as a decentralized combinatorial auction. Each host is treated as a self-interested agent with private valuations over bundles of defensive actions. These valuations are extracted empirically from the Q-values generated by reinforcement learning agents operating in the CAGE Challenge 2 environment \cite{kiely2023cage}. This allows our planner to incorporate host heterogeneity, action synergies, and limited resource budgets into a scalable, learning-based framework. Figure \ref{fig:pipeline} illustrates the end-to-end pipeline of our strategic cyber planning framework, from simulation-based valuation learning to auction-driven resource allocation.

Decentralization reflects the practical reality that hosts differ in roles, exposure, and impact and often operate under limited or local information. Combinatorial allocation is used because the utility of defensive actions is often non-additive: for instance, a host may benefit from \textit{Analyze} only if followed by \textit{Restore}, and some combinations may be redundant or conflicting. Allowing agents to bid on bundles captures these interdependencies more accurately than single-action models. We also aim for incentive compatibility, ensuring that agents (or their learned proxies) have no benefit from misreporting valuations. This improves the robustness of the allocation in learned systems, where preferences may be noisy or adaptive. Rather than maximizing total utility (as in VCG), we optimize for revenue, defined as the sum of reported valuations for assigned bundles, which aligns with truthful mechanism design under budget constraints and is tractable to learn with modern neural networks.

To solve the auction, we build on recent advances in differentiable mechanism design \cite{pham2025}. Specifically, we adopt CAFormer, a permutation-equivariant neural auction model that learns incentive-aligned resource allocations under feasibility constraints using adversarial regret minimization. In this work, we apply CAFormer to a new domain: upstream cyber defense planning under operational budgets.

We instantiate this approach in the DARPA CAGE Challenge 2 simulation environment, where private valuations for defensive actions are derived from reinforcement learning Q-values across heterogeneous hosts. We use CAFormer to allocate limited resources, such as Analyze or Restore actions, before execution begins, with the goal of improving strategic preparedness under threat. Our experimental results evaluate both the auction performance in terms of revenue and regret, and its downstream impact on cyber defense effectiveness.

The remainder of the paper is organized as follows. Section II reviews related work on cyber planning and mechanism design. Section III formulates the auction-based planning problem. Section IV describes our valuation modeling and environment setup. Section V details the CAFormer architecture and the training procedure. Section VI presents the experimental results. In Section VII, we conclude with a discussion of limitations and future directions.

\section{Related Work}

\subsection{Cybersecurity Planning: Centralized and Sequential Approaches}

Strategic planning in cyber defense has traditionally been approached through game-theoretic, graph-based, and decision-theoretic models. Stackelberg security games \cite{zhu2015game, alpcan2010network} model defender-attacker interactions in a leader-follower framework. These methods capture adversarial dynamics and resource constraints through mixed strategies, but often assume centralized control and perfect observability, limiting their scalability in realistic, distributed environments.

\subsection{Autonomous Cyber Defense and Gym Environments}

Vyas et al. \cite{vyas2023acd} provide a comprehensive survey on Automated Cyber Defense (ACD), formalizing core concepts and identifying key requirements for ACD agents and evaluation platforms. Their work emphasizes the importance of realistic Autonomous Cyber Operation (ACO) gyms, such as CyberBattleSim \cite{msft:cyberbattlesim}, CybORG \cite{cyborg_acd_2021}, FARLAND \cite{molina2021farland}, and EIReLaND \cite{cheung2023eireland}, which simulate adversarial cyber environments to support reinforcement learning-based defense strategies. The TTCP CAGE challenge series \cite{kiely2023cage} serves as a standardized testbed to benchmark ACD agents under realistic network dynamics and attack strategies. Studies such as Foley et al. \cite{foley2022mindrake} and Wolk et al. \cite{wolk2022beyondcage} propose hierarchical and ensemble-based PPO agents for CAGE Challenge 1 and 2. Kiely et al. \cite{kiely2023cage} further analyze these submissions, highlighting the trade-offs between DRL architectures, attacker randomization, and response strategies.

\subsection{Attack Graphs and Vulnerability Modeling}

Attack graphs provide a structured way to analyze system vulnerabilities and potential exploit paths \cite{sheyner2002automated, lippmann2005attackgraph, yun2011cvss}. Defense planning is typically cast as a graph optimization or search problem, where limited resources are allocated to protect the most vulnerable nodes. However, such approaches struggle to adapt to dynamic adversaries and do not naturally model action synergies or host-level heterogeneity.

\subsection{Decision-Theoretic Planning and Reinforcement Learning}

MDPs and their extensions, such as robust MDPs and POMDPs, have been applied to long-term cyber operations \cite{schlenker2017don, durkota2015optimal}. These models enable probabilistic reasoning under uncertainty, but often require hand-crafted reward functions and centralized planning. More recently, reinforcement learning has been used to learn tactical cyber defense behaviors in simulation \cite{nguyen2023survey, li2021survey}. Although effective for in-episode adaptation, these methods rarely address upstream resource allocation before an attack occurs.

\subsection{Auctions and Mechanism Design for Security}

Mechanism design provides a framework for allocating resources among agents with private preferences in a decentralized and incentive-compatible way. Although widely applied in economics and networking (e.g. spectrum auctions), its application to cybersecurity remains limited. Feng et al. \cite{zhang2018auction} use combinatorial auctions to allocate intrusion detection resources in cloud settings. Gupta et al. \cite{gupta2021} propose auction-based DoS defense. These approaches highlight the potential of market-based coordination in security but do not address upstream cyber planning or strategic pre-allocation of defensive actions.

\subsection{Differentiable Mechanism Design}

Recent work on differentiable mechanism design \cite{duetting2019optimal, rahme2021equivariantnet, ivanov2022regretformer, duan2024amenunet} has enabled the learning of auction mechanisms directly from data. These models optimize objectives such as revenue or regret while enforcing incentive compatibility through neural architectures. CAFormer \cite{pham2025} extends this paradigm by supporting permutation invariance, combinatorial feasibility, and adversarial regret training.

To our knowledge, this is the first application of neural auctions to upstream cyber defense planning. By combining learned valuations, obtained from simulation, with constraint-aware auction training, our work introduces a new avenue for scalable, interpretable, and decentralized cyber planning in adversarial environments.
\section{Problem Formulation}

We consider the upstream allocation of defensive actions from a finite action set $\mathcal{A}$ (e.g., \textit{Analyze}, \textit{Remove}, \textit{Restore}) to $N$ hosts before an attack episode. Each host $i \in \{1, \ldots, N\}$ has a private valuation function $v_i: 2^{\mathcal{A}} \rightarrow \mathbb{R}_+$ that assigns utility to the action bundles $S \subseteq \mathcal{A}$. These valuations, derived from reinforcement learning Q-values, represent expected defense impact conditioned on pre-attack system state.

Each host reports a bid vector $v_i = [v_i(S_1), \ldots, v_i(S_M)]$ over a predefined set of feasible bundles $\mathcal{S} = \{S_1, \ldots, S_M\}$. The planner selects an allocation distribution $\{x_{iS}\}$, where $x_{iS} \in [0,1]$ represents the probability of assigning bundle $S$ to host $i$, and payments $\{p_i\}_{i=1}^N$ that maximize total revenue:
\begin{align}
\max_{\{x_{iS}\}, \{p_i\}} \quad & \sum_i p_i \label{eq:revenue_objective} \\\
\text{s.t.} \quad 
& \text{Feasibility} \nonumber \\\
& \text{Incentive Compatibility} \nonumber \\\
& \text{Individual Rationality} \nonumber
\end{align}
where feasibility requires that each host receives at most one bundle, $\sum_S x_{iS} \leq 1$, and that each individual action is assigned to at most one host across all bundles, i.e., $\sum_{i} \sum_{S \ni a} x_{iS} \leq 1$ for all $a \in \mathcal{A}$. Incentive compatibility means that hosts have no incentive to misreport their valuations—ensured approximately by minimizing \textit{ex post regret} (defined in Section \ref {sec:auction_mechanism}), which measures how much better a host could have done by lying about its preferences. Individual rationality guarantees that each host’s expected utility is non-negative, $v_i - p_i \geq 0$, so that participation in the mechanism is voluntary.

\section{Valuation Generation}
We evaluate our method using the DARPA CAGE Challenge 2 environment—a high-fidelity cyber defense simulation built on the CybORG platform~\cite{kiely2023cage}. CC2 places a Blue team agent in charge of a 13-host network comprising user workstations, enterprise clients, defender nodes, and operational servers. We focus on Scenario~2, which simulates a manufacturing plant targeted by persistent phishing and lateral movement attacks. The Red team adversaries follow scripted policies such as \textit{BLine} and \textit{Meander}, while the Blue agent must maintain operational resilience over a fixed episode horizon (e.g., 30–100 time steps), by effectively responding to intrusions without disrupting critical assets.

We adopt the open-source top-performing Blue agent developed by Team CardiffUni \cite{cardiff2022cage2}, which won first place in the DARPA CAGE 2 challenge. This agent extends a standard Proximal Policy Optimization (PPO) backbone with several greedy domain-specific enhancements. This agent introduces host-specific decoy actions (nine in total), allowing more targeted deception strategies based on host identity and previous exploit traces. In addition, it augments the observation space with scan status memory vectors to track the history, allowing for informed sequencing of decoys and defensive actions. The agent also includes adaptive fallback behavior: if the host's decoy capacity is exhausted, it automatically executes a \textit{Remove} action, ensuring that decisions remain productive under constraints. These innovations lead to substantially improved performance over baseline agents and a realistic downstream policy for our upstream auction-based planner.

\subsection{Q-values Estimation}

Immediate reward signals in CC2—such as small positive values for \textit{Remove} or harsh penalties for \textit{Restore}—are often sparse, highly delayed and rarely updated. During training, the reward structure is dominated by long sequences of zero or negative values, leading to weak learning gradients and poor guidance for early strategic decisions. As a result, reinforcement learning agents trained directly on rewards tend to stagnate or overly exploit defensive actions with short-term gains. To address this, we instead use Q-values estimated from the agent's critic network, which reflect the long-term expected return from taking an action in a given state. These values capture both immediate consequences and downstream effects under adversarial dynamics. We compute $Q_i(s, a)$ using the standard advantage decomposition:
\[
Q_i(s, a) = V_i(s) + A_i(s, a),
\]
where $V_i(s)$ is the estimated state value and $A_i(s, a)$ is the advantage term. This formulation reduces variance and captures both the immediate and downstream consequences of decisions.

To ensure consistent scaling and tractable optimization, we normalize the input valuation matrix $X$ globally using min-max scaling:
\[
X = \frac{X - \min(X)}{\max(X) - \min(X) + \varepsilon},
\]
where $\varepsilon$ is a small constant to avoid division by zero. This transformation ensures that $X \in [0,1]$, stabilizing training and enabling gradient-based optimization for adversarial misreports, which are constrained to the same domain via projection (as in RegretNet \cite{duetting2019optimal}). Without normalization, the learned Q-values can be negative or poorly scaled, leading to invalid economic interpretations (e.g., negative payments) and difficulty in convergence. 
Since scaling is applied uniformly, it preserves the relative ranking of Q-values and maintains incentive properties.

\subsection{Valuation Model}
Each host $i$ defines a private valuation function $v_i(S)$ over bundles $S \subseteq \mathcal{A}$ of the core defensive actions. These include \textit{Analyze}, \textit{Remove}, and \textit{Restore}, which are consistently defined across hosts and directly comparable. 

Actions such as \textit{Deploy Decoy} and \textit{Activate Decoy} are excluded from the valuation model because they are dynamically generated at runtime based on host context (e.g., decoy availability, scan results) and do not correspond to static Q-value entries. Similarly, the \textit{Sleep} action is excluded due to its ambiguous utility and limited strategic relevance in offline planning.

In combinatorial auction settings, agents' valuations over bundles of items often exhibit interactions: the value of a bundle may not be simply the sum of its parts. We model such interactions using structured curvature terms in our valuation function.

We define the value of a bundle $S \subseteq \mathcal{A}$ to agent $i$ as:

$$
v_i(S) = \sum_{a \in S} Q_i(a) \pm \theta (|S|-1)^2 \cdot \epsilon_i(S)
$$
where $Q_i(a)$ is the per-action utility of action $a$ for agent $i$, estimated via a learned Q-function, $|S|$ is the size of the bundle, $\theta \in \{\lambda, \mu\}$ is a hyperparameter governing the strength of interaction (with sign depending on submodular or supermodular modeling), $\epsilon_i(S) \sim \mathcal{U}(0, 1)$ is a random multiplicative perturbation capturing preference uncertainty or misestimation.

We categorize the values of bundles into three types:
\begin{itemize}
\item \textbf{Additive:} $v_i(S) = \sum_{a \in S} Q_i(a)$. Assumes no interactions; each action contributes independently to the total value.
\item \textbf{Submodular:} $v_i(S) = \sum_{a \in S} Q_i(a) - \lambda (|S|-1)^2 \cdot \epsilon_i(S)$. Models diminishing returns—common in strategic decision-making—where additional actions are less valuable when similar actions are already selected.
\item \textbf{Supermodular:} $v_i(S) = \sum_{a \in S} Q_i(a) + \mu (|S|-1)^2 \cdot \epsilon_i(S)$. Captures synergistic behavior between actions, where joint use is more effective than isolated decisions.
\end{itemize}

The quadratic curvature term $|S|^2$ reflects the growing influence of action interactions as bundle size increases: In submodular settings, the cost of redundancy grows super-linearly: performing many similar actions leads to inefficient overuse of resources or overlapping effects; in supermodular settings, synergistic value compounds with larger bundles, e.g., joint restoration and analysis provides more insight than either action alone. This design approximates the number of pairwise interactions in a bundle, which scales as $\binom{|S|}{2} \sim \mathcal{O}(|S|^2)$. Thus, the quadratic form is a computationally efficient surrogate for explicitly modeling every pairwise action interaction.

Additionally, the PPO learned agents may exhibit uncertainty in their preferences due to partial observability, stochastic dynamics, or imperfect learning; and heterogeneity across hosts, even under similar conditions, especially in complex multi-agent environments like CC2. We incorporate a multiplicative random term $\epsilon_i(S)$ to reflect this uncertainty in the valuation curvature. This ensures that the curvature is stochastically variable across instances, making the auction training robust to host-level noise. The model can represent diverse host preferences while maintaining consistency in the base Q-value structure.

\section{Model Architecture and Training Procedure} \label{sec:auction_mechanism}

We adopt the CAFormer framework \cite{pham2025}, a differentiable combinatorial auction neural network based on Transformer layers. Given agent valuations $v_i(S)$ derived from Q-values (optionally modified via curvature), CAFormer predicts a \textit{probabilistic allocation} $z \in (0,1)^{n\times m}$ over feasible bundles and a \textit{normalized price vector} $\tilde{p} \in [0,1]^n$, indicating a fraction of the original bids, where $n$ is the number of host agents and $M = 2^{|\mathcal{A}|} - 1$ is the number of non-empty action bundles.

The model is trained to maximize total revenue, corresponding to minimizing the following loss:
\[
\mathcal{L}_{\text{revenue}} = -\sum_{i=1}^n \tilde{p}_i \cdot \sum_{S} z_{iS} \cdot v_i(S)
\]
where $v_i(S) \in [0,1]$ are normalized valuations. This objective is common in economic mechanism design, and its continuous structure supports stability and interpretability during training. Normalization preserves ordinal preferences while enabling bounded optimization.

To encourage incentive compatibility, we also define the \textit{regret} for agent~$i$ as the potential utility gain from misreporting:
\[
\text{Regret}_i = \max_{v'_i} \left[ u_i(v'_i, v_{-i}) - u_i(v_i, v_{-i}) \right]
\]
where $u_i$ denotes the utility of agent~$i$, defined as the expected value minus payment, and $v'_i$ represents a misreported valuation vector. The regret is estimated through a differentiable gradient ascent on $v'_i$, followed by projection into the feasible range $[0,1]^d$, as in \cite{duetting2019optimal}, which is included in the total loss:
\[
\mathcal{L} = (1 - \gamma) \log(1 + \mathcal{L}_{\text{revenue}}) + \gamma \mathcal{L}_{\text{regret}}
\]
where $\mathcal{L}_{\text{regret}} = \sum_{i=1}^n \text{Regret}_i$ and $\gamma$ controls the trade-off between revenue and incentive alignment.

In our setting, revenue represents the total value-weighted payment charged to agents for receiving defensive action bundles. Although maximizing total utility may appear to be more socially optimal, it implicitly assumes that resources are free and unconstrained. In practice, cyber defense involves operational costs, such as bandwidth, analyst time, or mission risk exposure, which must be taken into account in planning. The revenue objective reflects this trade-off, encouraging high-value allocations that are also cost-effective. Notably, our loss formulation remains flexible: It can be adapted to alternative goals (e.g., utility maximization, budget constraints, or fairness) without changing the core architecture.

\section{Experimental Results}

\subsection{Q-Value Distributions Across Host Types}

\begin{figure}[htbp]
    \centering
    \includegraphics[width=0.9\linewidth]{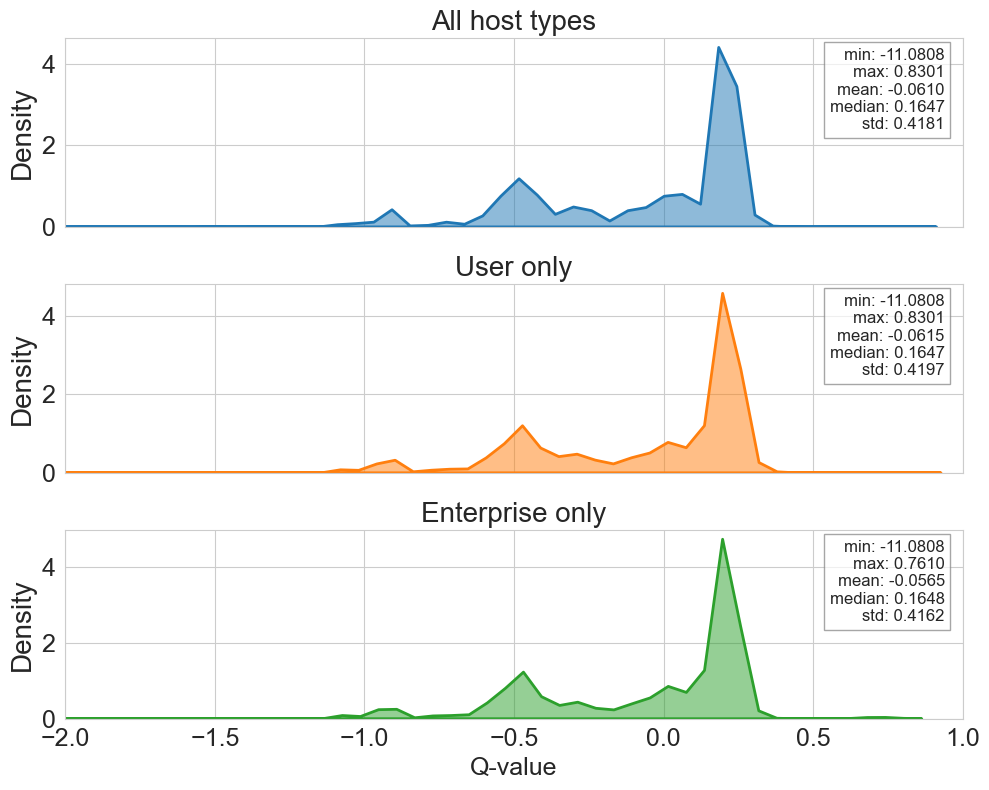}
    \caption{Distribution of Q-values for different host types. The \textit{All host types} setting includes users, enterprises, operator and defenders, while \textit{User only} and \textit{Enterprise only} isolate those roles to reveal type-specific patterns. Although the distributions are visually similar, Kolmogorov–Smirnov (KS) test reveals subtle significant differences between them ($\text{KS statistics} = 0.0695$ with $p < 0.001$). All distributions are heavily right-skewed and centered around mild positive values, with \textit{long negative tails} reaching as low as $-11$. This indicates that certain bundles or actions are \textit{strongly undesirable}, possibly due to redundancy or high risk. While visually similar, user Q-values show slightly higher variance, potentially reflecting \textit{more dynamic behavior patterns} compared to enterprise agents. The mixed setting exhibits broader density due to the inclusion of Defender and Operator, whose behavior involves probing, decoying, and deception, increasing heterogeneity.}
    \label{fig:qval_distributions}
\end{figure}

This paper introduces the use of Q-values as proxies for long-term strategic valuations of action-resource bundles in CC2, thus it is reasonable to question if Q-values capture hosts' preferences. Figure~\ref{fig:qval_distributions} shows the kernel density estimates (KDEs) of these values. 

These Q-value distributions reveal a stable structure across host types, validating their use as inputs for mechanism design. By quantifying long-term utilities directly from an RL-trained agent, we enable auction-based decision layers to inherit strategic foresight from lower-level learning systems. 

\subsection{Training Dynamics}
As a baseline, we implement an oracle allocation mechanism that maximizes social welfare by solving the winner-determination problem using Gurobi, which returns a deterministic allocation. This mechanism assumes access to Q-derived valuations and full bundle feasibility, enabling comparison against our learned model. For pricing, we compute VCG payments using the Clarke pivot rule. This mechanism is incentive-compatible under strategic misreports, but not revenue-maximizing, making direct comparisons with CAFormer less meaningful in performance terms. However, our focus is on understanding allocation behavior under truthful and strategic inputs. Benchmark comparisons of CAFormer are found in Pham et al. \cite{pham2025}.

\begin{figure}[htbp]
\centerline{\includegraphics[width=0.9\linewidth]{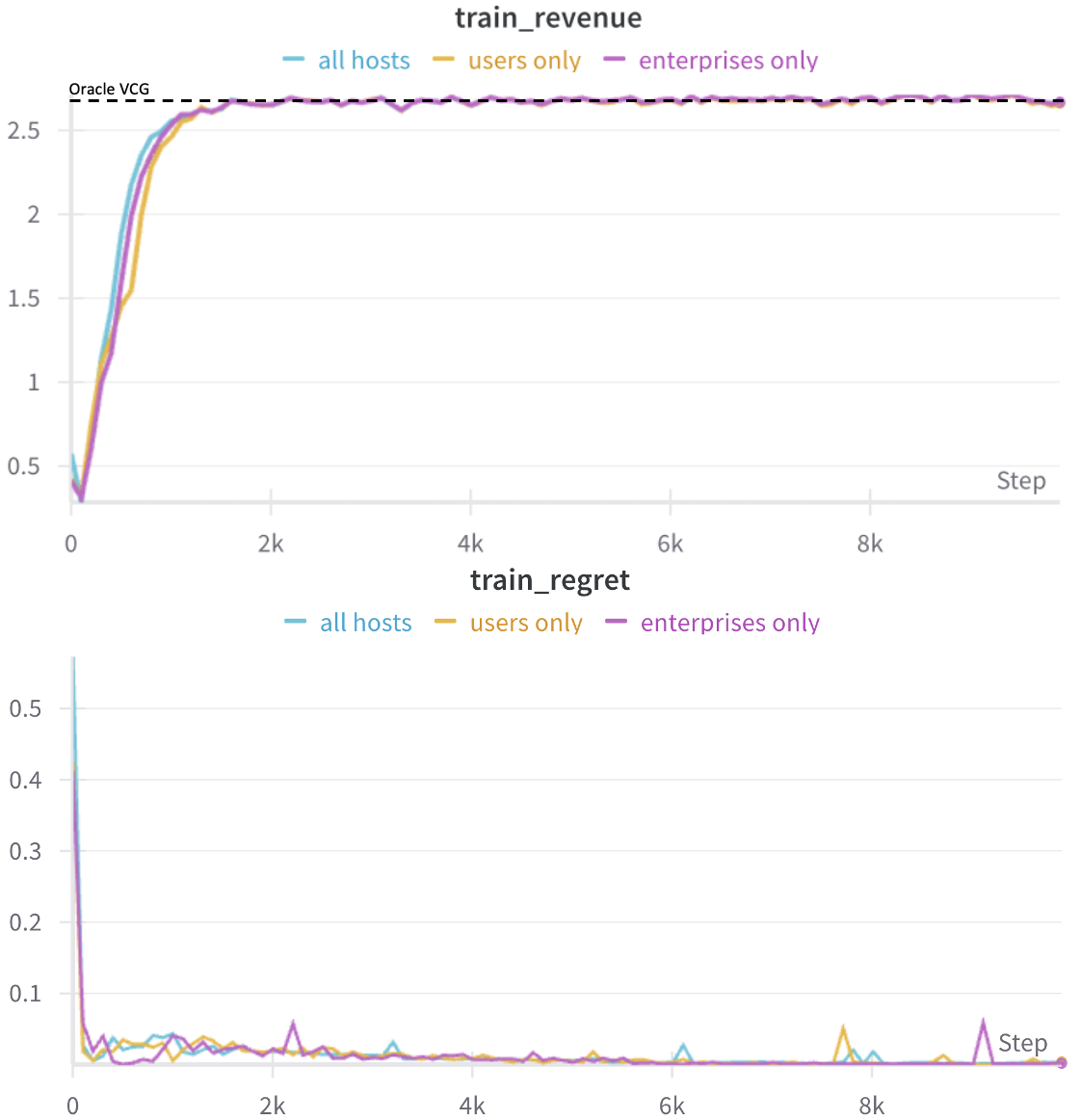}}
\caption{Training curves showing average revenue and regret for CAFormer and baseline oracle-VCG. Both revenue and regret converge effectively. The oracle-VCG gives 2.676 for all hosts, 2.6752 for enterprises only, and 2.677 for users only. Regrets are close to zero in 10,000 iterations. All experiments are averaged after 3 runs with standard deviation $\le$ 0.001}
\label{fig:training_plot}
\end{figure}

\subsection{Effect of Truthfulness on Allocation}
\begin{figure}[htbp]
\centerline{\includegraphics[width=0.9\linewidth]{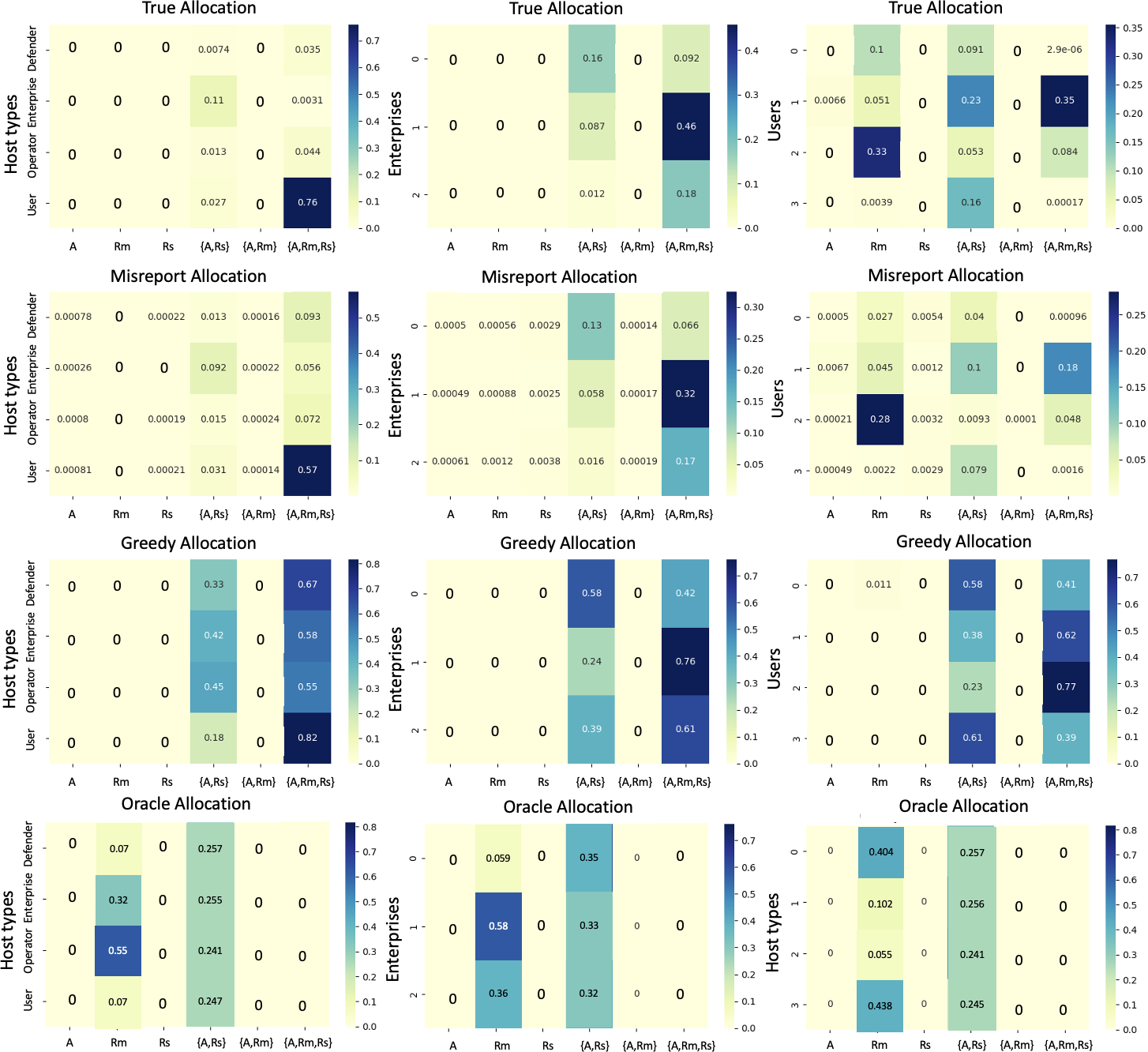}}
\caption{Allocation matrices under truthful vs. perturbed agent valuations. Under truthful reports, the learned mechanism concentrates most on \textit{Remove} and \{\textit{Analyse, Remove}\} actions, indicating a consistent preference for aggressive defense strategies. When agents misreport, allocations become more diffused across action bundles, suggesting that the mechanism is sensitive to manipulation. However, the relative importance of hosts (the allocation order) remains mostly preserved, demonstrating structural robustness. The greedy baseline over-allocates to the most comprehensive bundle \{\textit{Analyse, Remove, Restore}\}, particularly for user hosts, underscoring the inefficiency of naive valuation-based strategies (note that the rows of the greedy allocation heatmap do not sum to one, as each allocation represents the frequency of selecting a specific action across batches). This confirms that while our mechanism is not fully strategyproof, it maintains coherent allocation priorities and performs more adaptively than naive methods. While CAFormer and Greedy tends to prefer the comprehensive bundle, the oracle chooses bundle 1 (\textit{Remove}) more often because it can isolate the marginal value of just \textit{Remove} in some contexts. It does not favor the largest bundle as strongly, because it is often not strictly better than the sum of smaller bundles, especially in additive settings. }
\label{fig:allocation_matrix}
\end{figure}

We compare allocation behaviors under four conditions: truthful reporting, strategic misreporting, oracle, and greedy heuristic (Figure~\ref{fig:allocation_matrix}). The greedy allocation baseline assigns to each agent the single action bundle that yields the highest individual valuation, selecting the action with the highest bid per agent without considering global feasibility or incentive alignment. The oracle and greedy allocations are averaged across the samples. Interestingly, we observe that user-type hosts are frequently prioritized in learned allocation, often receiving aggressive actions such as \textit{Remove} and \{\textit{Analyse, Remove}\}. This may seem counterintuitive compared to static criticality rankings, where enterprises are often considered higher value. However, the Q-values driving our allocation reflect long-term strategic impact, suggesting that early disruption of User hosts may significantly hinder adversarial progress. This behavior aligns with the findings of the CC2 evaluation study \cite{claypoole2025interpret}, which shows that User hosts often serve as stepping stones toward more privileged targets. Thus, our mechanism implicitly learns to act preemptively, prioritizing early-stage containment.

.



\subsection{Alignment with Cyber Objectives}

To further assess whether the learned allocation mechanism prioritizes hosts involved in more cyber activity, we analyze the correlation between the aggregated allocation scores and the number of Red and Blue actions each host receives during the simulation. Allocation scores are aggregated by adding the model output in all bundles per host. We compute Pearson’s $r$ and Spearman’s $\rho$ correlation coefficients.

As shown in Table~\ref{tab:allocation_correlation}, we observe a strong positive correlation between allocation scores and the number of defensive actions by Blue in all hosts (Pearson $r=0.964,p=0.0361$), suggesting that the mechanism tends to focus attention on hosts where defenders are more active. The correlation with Red (attacker) activity is also positive, but weaker and not statistically significant. Within host-type subsets (e.g., Users or Enterprises), correlation trends are directionally similar, though only Blue (enterprises) reaches significance, likely due to the reduced sample sizes. These results indicate that, while the mechanism is not explicitly aware of the underlying mission-criticality metadata, its learned allocations align closely with observed operational activity in the environment.

\begin{table}[ht]
\centering
\caption{Correlation between allocation scores and number of red/blue actions across hosts.}
\begin{tabularx}{\linewidth}{ccc}
\toprule
\textbf{Correlation} & \textbf{Pearson's $r$ (p)} & \textbf{Spearman's $\rho$ (p)} \\
\midrule
Red Agent (all)         & 0.529 ($p=0.47$) & 0.800 ($p=0.20$) \\
Blue Agent (all)        & \textbf{0.964} ($p=0.04$) & 0.800 ($p=0.20$) \\
Red Agent (enterprises) & 0.351 ($p=0.77$) & 0.500 ($p=0.67$) \\
Blue Agent (enterprises)& 0.753 ($p=0.46$) & \textbf{1.000} ($p=0.00$) \\
Red Agent (users)       & 0.845 ($p=0.15$) & 0.800 ($p=0.20$) \\
Blue Agent (users)      & 0.769 ($p=0.23$) & 0.600 ($p=0.40$) \\
\bottomrule
\end{tabularx}
\label{tab:allocation_correlation}
\end{table}

\section{Conclusion}

We present a new approach to cyber operation planning that models the allocation of defensive actions as a combinatorial auction, with valuations derived from RL-based Q-values. These values capture the long-term impact of defensive interventions across heterogeneous host types. Using CAFormer, a neural auction architecture, we learn mechanisms that are incentive-aware and flexible to curvature in agent preferences. Compared to oracle and greedy baselines, our method yields robust and interpretable outcomes, even under strategic misreporting. Our approach leverages centralized Q-values as proxies for decentralized agent valuations, enabling tractable long-term planning. However, Q-values require a fully trained agent in the downstream task, which may not be available in other deployment settings, and introduces a modeling assumption: centralized Q-values reflect the global return under a shared policy rather than private agent utilities. Exploring training in host-decentralized environments could better bridge this gap. The framework offers a modular way to incorporate more principled cyber metrics (e.g., resilience scores) and opens directions for combining strategic planning with explainable decision support in adversarial environments.

\newpage

\section*{Acknowledgment}

This research was funded by the Defense Advanced Research Projects Agency (DARPA), under
contract W912CG23C0031.

\bibliographystyle{IEEEtran}
\bibliography{reference}

\end{document}